\documentstyle[twocolumn,epsfig,prl,aps]{revtex}
\begin{document}

\setlength{\topmargin}{-2.0cm}      % hep-ph 
%\setlength{\topmargin}{-2cm}

%%%%%%%%%%%%%%%%%%%%%%%%%%%%%%%%%%%%%%%%%%%%%%%%%%%
\sloppy
\newcommand{\lapprox}{%
\mathrel{%
\setbox0=\hbox{$<$}
%\setbox1=\hbox{$\sim$}
\raise0.6ex\copy0\kern-\wd0
\lower0.65ex\hbox{$\sim$}
}}
\newcommand{\gapprox}{%
\mathrel{%
\setbox0=\hbox{$>$}
%\setbox1=\hbox{$\sim$}
\raise0.6ex\copy0\kern-\wd0
\lower0.65ex\hbox{$\sim$}
}}
\newcommand{\be}{\begin{equation}}
\newcommand{\ee}{\end{equation}}
\newcommand{\bea}{\begin{eqnarray}}
\newcommand{\eea}{\end{eqnarray}}
\newcommand{\lbl}[1]{\label{eq:#1}}
\newcommand{ \rf}[1]{(\ref{eq:#1})}
\newcommand{\FF}{{\cal F}_{\pi^0\gamma^*\gamma^*}}

%------------------------------------------------------------------
\begin{titlepage}
\begin{flushright}
November 6, 2001 \\ 
Revised: January 31, 2002 \\ 
CPT-2001/P.4260
\end{flushright}

\vspace{1cm} 
\begin{center}
\begin{bf}

{\Large \bf Hadronic Light-By-Light Scattering Contribution to the
Muon $g-2$:\\[0.3cm] 
 An Effective F\/ield Theory Approach} \\[2cm]
\end{bf}
{\large M. Knecht, A. Nyf\/feler, M. Perrottet, and E. de Rafael}
\\[1cm]
 Centre de Physique Th{\'e}orique, CNRS--Luminy, Case 907\\ 
F-13288 Marseille Cedex 9, France \\[3cm]  

\end{center}

%\vspace*{3cm} 
%\vfill
\begin{center} {\bf Abstract }
\end{center} 
\begin{abstract}
The hadronic light-by-light contribution to $a_{\mu}$, the anomalous
magnetic moment of the muon, is discussed from the point of view of an
effective low-energy theory.  As an application, the coefficient of
the leading logarithm arising from the two-loop graphs involving two
anomalous vertices is computed, and found to be positive. This 
corresponds to a positive sign for the pion-pole contribution to the
hadronic light-by-light correction to $a_{\mu}$, and to a sizeable
reduction of the discrepancy between the present experimental value of
$a_{\mu}$ and its theoretical counterpart in the standard model.
\end{abstract}
%\end{center}
%\vfill

\vspace*{3cm} 

\begin{center} 
%\begin{minipage}[t]{12cm}
%{\bf PACS}:  13.40.Em, 12.20.Ds, 12.39.Fe, 14.60.Ef~\\[0.2mm]
%{\bf Keywords:} \begin{minipage}[t]{9.5cm}Muon magnetic moment,
%Effective field theory, Renormalization group, Chiral symmetry
%\end{minipage}
%\end{minipage} 

\hspace*{-0.95cm}{\bf PACS:} 13.40.Em, 12.20.Ds, 12.39.Fe, 14.60.Ef~\\[0.2mm]
{\bf Keywords:} \begin{minipage}[t]{7.5cm} Muon magnetic moment,
Effective field theory, Renormalization group, Chiral symmetry.
\end{minipage}

\end{center}

\end{titlepage}

\begin{titlepage}
${  }$
\end{titlepage}

%%%%%%%%%%%%%%%%%%%%%%%%%%%%%%%%%%%%%%%%%%%%%%%

\wideabs{
\title{Hadronic Light-By-Light Scattering Contribution to the Muon
$g-2$:\\[0.3cm] 
 An Effective F\/ield Theory Approach }
\author{
M. Knecht, A. Nyf\/feler, M. Perrottet, and E. de Rafael
}

\address{
Centre de Physique Th{\'e}orique, CNRS--Luminy, Case 907,
F-13288 Marseille Cedex 9, France}
\maketitle

\begin{abstract}
The hadronic light-by-light contribution to $a_{\mu}$, the anomalous
magnetic moment of the muon, is discussed from the point of view of an
effective low-energy theory.  As an application, the coefficient of
the leading logarithm arising from the two-loop graphs involving two
anomalous vertices is computed, and found to be positive.  This 
corresponds to a positive sign for the pion-pole contribution to the
hadronic light-by-light correction to $a_{\mu}$, and to a sizeable
reduction of the discrepancy between the present experimental value of
$a_{\mu}$ and its theoretical counterpart in the standard model.
\end{abstract}

}

The Brookhaven E821 experiment has recently measured
\cite{BNL2} the anomalous magnetic moment $a_{\mu}$ of the muon with a
precision of $\pm 16\times 10^{-10}$, improving by a factor of 6 on
the previous measurement at CERN \cite{cern77}.  As the full set of
data will be analyzed, the experimental error bars are expected to
decrease further, by at least an additional factor of 3.  The
discrepancy between the experimental value of Ref. \cite{BNL2} and
certain theoretical estimates can be as large as 2.6$\sigma$. Many
possibilities to explain this difference through the introduction of
new, beyond the standard model, degrees of freedom have therefore been
considered. It is however important to keep in mind that the
theoretical estimates of $a_{\mu}$ include several contributions
involving the nonperturbative hadronic sector of the standard
model. In this Letter, we shall focus on one of these contributions,
the so-called hadronic light-by-light scattering, which arises from
the lowest order contribution, in the fine-structure constant
$\alpha$, to the matrix element ($e$ stands for the electron charge)
\be 
\langle \mu^-(p\,')\vert (ie) j_{\rho}(0) \vert \mu^-(p) \rangle
\equiv 
(-ie) {\bar{\mbox{u}}}(p\,'){\widehat\Gamma}_{\rho}(p\,',p)\mbox{u}(p)
\ee
of the conserved light quark electromagnetic current,
$j_{\rho}=(2{\bar u}\gamma_{\rho}u - {\bar d}\gamma_{\rho}d - {\bar
s}\gamma_{\rho}s)/3$.  This matrix element is given by a two-loop
integral involving the connected fourth rank hadronic vacuum
polarization tensor $\Pi_{\mu\nu\lambda\rho} (q_1,q_2,q_3)$.
The corresponding contribution $a_{\mu}^{\mbox{\tiny{LbyL; had}}}$
to $a_{\mu}$ is equal to
\be
%a_{\mu}^{\mbox{\tiny{LbyL; had}}} = 
\lim_{p'-p\to 0}
{\mbox{tr}}[(\not\! p + m)\Lambda^{(2)}_{\rho}(\not\! p\,'+ m)
{\widehat\Gamma}^{\rho}(p\,',p)]\,,
\lbl{F2proj}
\ee
with ($k=p\,'-p$)
$$%\be
\Lambda^{(2)}_{\rho} = \frac{m^2}{k^2(4m^2-k^2)}\,\bigg[\gamma_{\rho}+
\frac{k^2+2m^2}{m(k^2-4m^2)}(p\,'+p)_{\rho}\bigg].
%\lbl{proj2}
$$%\ee

This hadronic light-by-light correction to $a_{\mu}$ has been studied
by several authors in the past
\cite{calmet76,KNO_85,HKS_95_96,BPP,persson01,bartos01}. 
Its value, and even its {\it
sign}, has suffered several changes, but the latest evaluations came
to values that agreed within the quoted theoretical error bars and
that were {\it negative}.  In particular, there exists a well defined
contribution to $a_{\mu}^{\mbox{\tiny{LbyL; had}}}$, denoted by
$a_{\mu}^{\mbox{\tiny{LbyL;}}\pi^0}$, that arises upon restricting
$\Pi_{\mu\nu\lambda\rho} (q_1,q_2,q_3)$ to its reducible one-pion
exchange component and which represents about 70\% of the total value
of $a_{\mu}^{\mbox{\tiny{LbyL; had}}}$.  Very recently, two of us,
reconsidering this evaluation of $a_{\mu}^{\mbox{\tiny{LbyL;}}\pi^0}$,
found a {\it positive} result \cite{KN_01}, but which, in absolute
value, agreed with the previous numerical values whenever comparison
was possible. Resolving this sign ambiguity certainly represents a
major issue, since the result of \cite{KN_01} reduces the theory
vs experiment discrepancy to less than 1.5$\sigma$. This Letter aims
at providing an argument in favor of a positive sign for
$a_{\mu}^{\mbox{\tiny{LbyL;}}\pi^0}$ which, to a large extent, does
not rely on the methodology followed in Ref. \cite{KN_01}. The
off-shell pion-photon-photon form factor $\FF$ constitutes an
important ingredient for the evaluation of
$a_{\mu}^{\mbox{\tiny{LbyL;}}\pi^0}$.  For a constant form factor,
fixed by the Wess-Zumino-Witten (WZW) term of QCD~\cite{WZW}, which
reproduces the Adler~\cite{adler69}, Bell-Jackiw~\cite{belljackiw} 
anomaly, the corresponding two-loop integral for
$a_{\mu}^{\mbox{\tiny{LbyL;}}\pi^0}$ diverges like
$(\alpha/\pi)^3{\cal C}\ln^2\Lambda$, where $\Lambda$ stands for an
ultraviolet cutoff \cite{melnikov01}.  We shall be interested in the
determination of the coefficient ${\cal C}$ of this log-squared
divergence, so that we may then compare it to the value that can be
extracted from the analysis of Ref. \cite{KN_01}. In order to achieve
this goal, we shall use a renormalization group argument, along the
lines discussed in Ref. \cite{weinberg79}, within the framework of the
effective low-energy field theory of the standard model.  Before doing
so, we first describe the relevant features of this effective theory.

\vspace*{0.1cm} 
The low-energy degrees of freedom of the standard model
involve the pseudoscalar mesons, the light leptons, and the
photon. The interactions between these degrees of freedom are
constrained by the symmetries of the standard model, like chiral
symmetry or $U(1)$ gauge invariance.  This effective theory will
provide a good description of physical processes as long as the energy
scales involved are much lower than a typical hadronic scale
$\Lambda_H\sim 1$~GeV. Since the leptons obey a first order equation
of motion, the chiral counting has to be suitably adapted. The
situation differs from the case of chiral perturbation theory in the
presence of baryons due to the fact that the mass $m$ of the light
leptons does not introduce a scale of the order of $\Lambda_H$.
Rather, the counting $m\sim{\cal O}({\mbox{p}})$ arises quite
naturally.  Furthermore, we shall also count fermion bilinears as
quantities of order ${\mbox{p}}$, as is the case in the sector of the
effective theory which describes the semileptonic decays of the
pseudoscalar mesons \cite{knecht_00} (see also Ref. \cite{nyffeler}
for a discussion of this aspect in a different context).  Finally, the
electric charge is also counted as a quantity of order ${\mbox{p}}$
\cite{urech95}. The lowest order term in this extended chiral
expansion then starts at order ${\mbox{p}}^2$, with (for the notation,
see Refs. \cite{gassleut85,ecker89,urech95}):
\bea
{\cal L}^{(2)} &=& -\,\frac{1}{4}\,F_{\mu\nu}F^{\mu\nu} 
+ {\overline\psi}(i\not\!\!D - m)\psi
+ e^2C\langle QU^+QU \rangle  
\nonumber\\
&&\!\!\!\!\!\!
+ \frac{F_0^2}{4}\,\big(
\langle d^{\mu}U^+d_{\mu}U \rangle + 2B_0\langle {\cal M}^+U +
U^+{\cal M} \rangle \big) 
\,.
\eea
This lowest order Lagrangian involves the Maxwell Lagrangian for the
photon, the tree-level minimal coupling of leptons and photons, as
well as the lowest order chiral Lagrangian for the mesons, together
with their couplings to the photons.  It gives the (finite) lowest
order, ${\cal O}({\mbox{p}}^6)$, contribution to
$a_{\mu}^{\mbox{\tiny{LbyL; had}}}$ obtained upon restricting
$\Pi_{\mu\nu\lambda\rho} (q_1,q_2,q_3)$ to its lowest order
approximation, consisting of a charged pion loop with pointlike
electromagnetic vertices.  At ${\cal O}({\mbox{p}}^4)$ in the
Lagrangian, we find the usual counterterms involving the low-energy
constants $L_i$ \cite{gassleut85}, $K_i$
\cite{urech95,neufeld9596} and $X_i$ \cite{knecht_00}. 
These counterterms absorb the divergences
due to loops with virtual pions, photons, and leptons.  For instance, $L_9$ is
needed in order to make the slope of the pion form factor, generated
by a charged pion loop, finite.  Note that the pion loop contribution
to $\Pi_{\mu\nu\lambda\rho} (q_1,q_2,q_3)$ with these nonpointlike
vertices, gives a next-to-leading, ${\cal O}({\mbox{p}}^8)$,
correction to $a_{\mu}^{\mbox{\tiny{LbyL; had}}}$ which is no longer
finite, and thus requires a corresponding counterterm from ${\cal
L}^{(10)}$.  In addition, we also encounter in ${\cal L}^{(4)}$ the
counterterms which result from the divergent loop graphs involving
virtual leptons and photons, such as  the ${\cal O}(\alpha)$
renormalizations of $e$, $m$ and of the lepton wave function. These
will play no role in the present discussion.  Finally, the WZW
Lagrangian of the odd intrinsic parity sector also occurs in ${\cal
L}^{(4)}$.  For our present purpose, we need only to retain a piece
of the latter (our conventions are as follows:
$\gamma_5=i\gamma^0\gamma^1\gamma^2\gamma^3$, $\varepsilon_{0123}=+1$) 
\bea
{\cal L}^{(4)} &=& \frac{ie^2N_C}{24\pi^2}\varepsilon_{\mu\nu\alpha\beta}
\partial^{\mu}A^{\nu}A^{\alpha}\langle Q^2\partial^{\beta}UU^+
+ Q^2U^+\partial^{\beta}U
\nonumber\\
&&
 - \frac{1}{2}QUQ\partial^{\beta}U^+
+ \frac{1}{2}QU^+Q\partial^{\beta}U \rangle
+ \ldots , \nonumber \\
& = & - {\alpha N_C \over 12 \pi F_0}
\varepsilon_{\mu\nu\alpha\beta} F^{\mu\nu} A^{\alpha}\partial^{\beta} 
\pi^0 + \ldots  \,. \lbl{L4WZW}
\eea
This term is responsible for the lowest order, and divergent,
contribution to the pion-pole correction 
$a_{\mu}^{\mbox{\tiny{LbyL;}}\pi^0}$, which thus starts at order
${\mbox{p}}^{8}$ and corresponds to the diagrams (a), (b), and
(c) of Fig.~1.  At order ${\mbox{p}}^6$ in the effective field
theory, we encounter, for instance, divergent loops involving the
particular WZW vertex from ${\cal L}^{(4)}$ shown in Eq. \rf{L4WZW}
and a virtual fermion line, like precisely the triangular subgraphs in
the two first graphs of Fig.~1.  These divergences have to be
canceled by an appropriate set of counterterms which, according to
Ref.~\cite{savage92}, reads
\bea
{\cal L}^{(6)} &=& \frac{3i\alpha^2}{32\pi^2}\,
{\overline\psi}\gamma_{\mu}\gamma_5\psi\bigg\{
\chi_1\langle Q^2(U^+d^{\mu}U + d^{\mu}UU^+)\rangle 
\nonumber\\
&&
+ \chi_2 \langle QU^+Qd^{\mu}U - Qd^{\mu}U^+QU\rangle \bigg\} 
+ \ldots  \, . 
\lbl{L6chi}
\eea
Here also, there may be other terms, indicated by the ellipsis, but
which are of no relevance for the discussion concerning
$a_{\mu}^{\mbox{\tiny{LbyL;}}\pi^0}$. Under renormalization, the terms
we have shown absorb the divergence of the triangular subgraphs in the
diagrams (a) and (b) of Fig.~1.  They therefore also
contribute to $a_{\mu}^{\mbox{\tiny{LbyL;}}\pi^0}$ at order
${\mbox{p}}^{8}$, through the diagrams (d) and (e) of
Fig.~1. We shall come back to this important issue below.  Let us
first conclude this brief description of the effective theory by
mentioning that the graphs (a) and (b) also have an overall
divergence, which is absorbed by an appropriate set of counterterms in
${\cal L}^{(10)}$, the effective Lagrangian at the ${\cal
O}({\mbox{p}}^{10})$ level.  The corresponding tree-level
contribution to $a_{\mu}^{\mbox{\tiny{LbyL;}}\pi^0}$ is represented by
the diagram (f) in Fig.~1. The detailed structure of ${\cal
L}^{(10)}$ will, however, not be needed.

%%%%%%%%%%%%%%%%%%%%%%%%%%%%%%%%%%%%%%%%%%
\begin{figure}[!t]%[!h]
\centerline{\psfig{figure=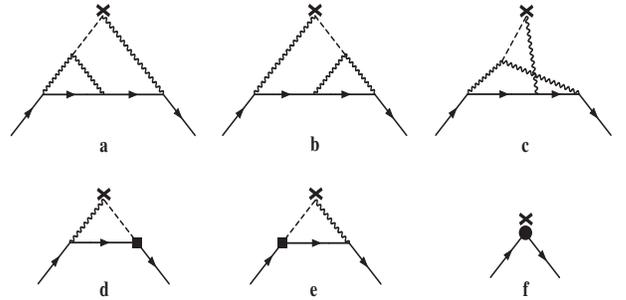,height=4cm,width=8cm}}
\caption{The graphs contributing to
$a_{\mu}^{\mbox{\tiny{LbyL;}}\pi^0}$ at lowest order in the effective
field theory. The two-loop graphs (a), (b), and (c) contain
two WZW vertices from \protect\rf{L4WZW}. The one-loop graphs (d)
and (e) have an insertion of one vertex
(\protect\rule{1.5mm}{1.5mm}) of ${\cal L}^{(6)}$,
Eq. \protect\rf{L6chi}. Finally, the tree-level graph (f) stems
from the ${\cal O}$(p$^{10}$) counterterms. Note that the diagram (c)
is actually finite.  } 
\label{fig:fig1}
\end{figure}
%%%%%%%%%%%%%%%%%%%%%%%%%%%%%%%%%%%%%%%%%

%\vspace*{0.4cm} 
We next return to the original purpose of this Letter, the
determination of the two-loop leading logarithm arising from the sum
of the diagrams (a) and (b) of Fig.~1. Since the diagram (c) is
finite, it is of no relevance for the present discussion. In the  
context of a renormalizable field theory, such a question would be
most naturally addressed by means of the renormalization group
equations. As stressed by Weinberg \cite{weinberg79}, the
renormalization group remains a useful concept even in the case of a
nonrenormalizable local effective field theory.  Within the effective
theory framework, the expression of the renormalized contribution to
$a_{\mu}$ arising from the graphs of Fig.~1 takes the general form
\be \label{a_mu_EFT}
%{\widehat F}_2^{(\pi^0)}(0)
a_{\mu}^{\mbox{\tiny{LbyL;}}\pi^0} = 
H\left(\frac{m}{\mu}\right)
+ \chi(\mu)J\left(\frac{m}{\mu}\right)
+ \kappa(\mu)\,.
\ee
Here $\mu$ denotes the arbitrary subtraction scale introduced by the
renormalization procedure. The contributions of the two-loop graphs
(a), (b), and (c) of Fig.~1 are given by the first term in
this expression. The next term describes the contributions from the
one-loop graphs (d) and (e), with an insertion of the
renormalized ${\cal O}({\mbox{p}}^6)$ counterterm $\chi\equiv
(-1/4)(\chi_1 + \chi_2)$, while the last term, $\kappa(\mu)$,
collectively stands for the renormalized tree-level contributions from
the ${\cal O}({\mbox{p}}^{10})$ effective Lagrangian. The dependence
on the subtraction scale $\mu$ in the two- and one-loop functions
reads
\bea
H\left(\frac{m}{\mu}\right)
&=&
\sum_{p=0,1,2}
h_p
\ln^p\left(\frac{m}{\mu}\right),
\nonumber\\
J\left(\frac{m}{\mu}\right)
&=&
\sum_{q=0,1}
j_q
\ln^q\left(\frac{m}{\mu}\right).
\eea
The (dimensionless) coefficients $h_p$ and $j_q$ do not depend on
$\mu$, but are functions of the ratios $M_{\pi^0}/m$ and $F_{\pi}/m$.
They therefore satisfy ${\cal D}h_p=0$, ${\cal D}j_q=0$, where
\be
{\cal D} \equiv m\frac{\partial}{\partial m} +
M_{\pi^0}\frac{\partial}{\partial M_{\pi^0}} +
F_{\pi}\frac{\partial}{\partial F_{\pi}}.
\ee
Being a physical quantity, $a_{\mu}^{\mbox{\tiny{LbyL;}}\pi^0}$ 
obeys the condition $\mu (da_{\mu}^{\mbox{\tiny{LbyL;}}\pi^0}/d\mu)=0$,
which gives
\bea
&&
\ln\left(\frac{m}{\mu}\right)\bigg[
2 h_2 -
\gamma_{\chi}j_1
\bigg] + h_1
- \gamma_{\chi}j_0
+ \chi(\mu) j_1
\nonumber\\
&&
-\mu\,\frac{d\kappa(\mu)}{d\mu}
= 0,
\eea 
where we have introduced $\gamma_{\chi} = \mu [d \chi(\mu) / d\mu]$.
Acting with the operator ${\cal D}$ on this equation and using the
fact that ${\cal D}\chi(\mu) = 0, {\cal D} \kappa(\mu) = 0$, one finds
that the term proportional to $\ln(m/\mu)$ has to vanish separately,
i.e., 
\be \label{h_2} 
h_2\left(\frac{M_{\pi^0}}{m},\frac{F_{\pi}}{m}\right) = 
{1\over 2}\gamma_{\chi}j_1\left(\frac{M_{\pi^0}}{m},\frac{F_{\pi}}{m}\right),
\ee
and that 
\be
\kappa(\mu) = 
{1\over 2}\gamma_{\chi}j_1\left(\frac{M_{\pi^0}}{m},\frac{F_{\pi}}{m}\right)
\ln^2\left({\mu\over\mu_0}\right) + \cdots
\,.
\ee
Therefore, in order to compute $h_2\equiv (\alpha/\pi)^3{\cal C}$, we
have to extract the dependence on $\ln(m/\mu)$ from the one-loop
graphs of Fig.~1.  This is our next step.

\vspace*{0.2cm} 
The sum of the two graphs (d) and (e) of Fig.~1 reads
\bea
&&
(-ie){\bar{\mbox{u}}}(p\,')
{\widehat\Gamma}_{\rho}^{\mbox{\tiny{(d)+(e)}}}(p\,',p)
{\mbox{u}}(p) =  \left(\frac{\alpha^2}{4\pi^2 F_\pi}\right)
\left(\frac{i\alpha N_C}{3\pi F_\pi}\right)
\nonumber\\
&&
\times
\chi(\mu)
\int \frac{d^4q}{(2\pi)^4}\, \frac{-i}{(k-q)^2}\,
\frac{i}{q^2-M_{\pi^0}^2}\,
\varepsilon_{\rho\lambda\alpha\beta}q^{\alpha}k^{\beta}  
\nonumber\\
&&
\times
{\bar{\mbox{u}}}(p\,')
\bigg[
(-ie)\gamma^{\lambda}\frac{i}{\not\!p\,+ \not\!q -m}\,\not\!q \gamma_5
\nonumber\\
&&
\qquad
+\, \not\!q \gamma_5\frac{i}{\not\!p\,' - \not\!q -m}(-ie)\gamma^{\lambda}
\bigg] {\mbox{u}}(p).
\eea
This integral diverges logarithmically. Introducing a cutoff $\Lambda$,
the corresponding contribution to $a_\mu$ is obtained upon using  
\rf{F2proj}. The Dirac trace can easily be performed, leading to the
result 
$a_{\mu}^{\mbox{\tiny{LbyL;}}\pi^0}\vert_{\mbox{\tiny{(d)+(e)}}}
=j_1\ln(m/\Lambda)+$finite, with
\be
j_1 = \frac{N_C}{24\pi^2}\,\left(\frac{\alpha}{\pi}\right)^3
\,\left(\frac{m}{F_{\pi}}\right)^2.
\ee
It is interesting to note that the result for  $j_1$  can also be obtained
from the calculation of the coefficient of the $\ln(M_Z^2/m^2)$ term in
the two-loop electroweak contribution to $a_{\mu}$; see
Ref. \cite{PPdR95}.  Next, we recall that the scale dependence of the
constant $\chi(\mu)$ is already known \cite{savage92,ametller93}.
Indeed, the same combination of $\chi_1$ and $\chi_2$ arises in the
${\cal O}(\alpha^2)$ contribution to the $\pi^0\to e^+e^-$ amplitude,
together with the divergent one-loop graph, generated by the WZW
vertex, and involving two virtual photons and one fermion line [the
on-shell restriction of the triangular subgraphs in diagrams (a)
and (b)].  The result reads $\gamma_{\chi}=N_C$, and thus with 
Eqs. (\ref{a_mu_EFT}) and (\ref{h_2}) we obtain 
\be
a_{\mu}^{\mbox{\tiny{LbyL;}}\pi^0} = \left( {\alpha \over \pi}
\right)^3 \left[ {\cal C} \ln^2\left( {m \over \mu_{0}} \right) + {\cal
O}\left[ \ln ( m / \mu_0) \right] \right],   
\ee
where 
\be
{\cal C} =
+3\,\left(\frac{N_C}{12\pi}\right)^2\,\left(\frac{m}{F_{\pi}}\right)^2.
\label{Clog2}
\ee
For $F_{\pi}=92.4$ MeV, $m=105.66$ MeV and $N_C=3$, this gives ${\cal
C}=0.0248$.  We shall now discuss some consequences of this result as
far as the sign of $a_{\mu}^{\mbox{\tiny{LbyL;}}\pi^0}$ 
obtained in Ref. \cite{KN_01} is concerned.

%\vspace*{0.4cm} 
Our analysis thus tells us that the sum of the graphs (a)
and (b) of Fig.~1 not only diverges, but behaves like
$(\alpha/\pi)^3{\cal C}\ln^2\Lambda$ as the ultraviolet cutoff
$\Lambda$ is sent to infinity. Actual calculations of
$a_{\mu}^{\mbox{\tiny{LbyL;}}\pi^0}$ [which, in the framework of the
effective theory, amount to estimates of $\chi(\mu)$ and of
$\kappa(\mu)$] replace the pointlike WZW vertex by a form factor
$\FF$ which regularizes the ultraviolet behavior. The models of $\FF$
that are introduced for this purpose in one way or another involve a
hadronic scale $\Lambda_H$, typically the $\rho$-meson mass $M_V$ (or
$\Lambda_H \sim 2 m_Q$, with a constituent quark mass $m_Q$.). As
$M_V\to\infty$, the form factor becomes constant, and one therefore
has to recover the same behavior,
$a_{\mu}^{\mbox{\tiny{LbyL;}}\pi^0}\sim (\alpha/\pi)^3{\cal C}\ln^2
M_V$.  To the best of our knowledge, this aspect of
$a_{\mu}^{\mbox{\tiny{LbyL;}}\pi^0}$ has not been discussed in detail
before. In particular, we know of no previous determination of the
constant ${\cal C}$. There is a reference to the $\ln^2\Lambda$
behavior in \cite{melnikov01}, which, in our notation, states that
$a_{\mu}^{\mbox{\tiny{LbyL;}}\pi^0}\vert_{WZW} \sim (\alpha/\pi)^3(m/4\pi
F_\pi)^2(\ln^2\Lambda)/2$, but it is not clear from the context of
Ref. \cite{melnikov01} whether the {\it sign} was meant to be part of
the estimate or not.

In Ref. \cite{KN_01}, $a_{\mu}^{\mbox{\tiny{LbyL;}}\pi^0}$ was expressed, 
for a certain class of form factors, in terms of a two-dimensional
integral representation of the form (we omit here the small
contribution from the finite diagram (c) in Fig.~1, which
plays no role in the present discussion) 
\bea
&&
a_{\mu}^{\mbox{\tiny{LbyL;}}\pi^0} = 
\left(\frac{\alpha}{\pi}\right)^3\!\!\int_0^{\infty}\!\!\!\!dQ_1\!\!
\int_0^{\infty}\!\!\!\!dQ_2\, [f(M_V\!,\!Q_1,\!Q_2) w_{f}(Q_1,\!Q_2) 
\nonumber\\
&&
\qquad\qquad\qquad\qquad + g(M_V\!,\!Q_1,\!Q_2) w_{g}(M_V\!,\!Q_1,\!Q_2) ] .
\label{2dInt}
\eea
The functions $f(M_V,Q_1,Q_2)$ and $g(M_V,Q_1,Q_2)$ depend
quadratically on $\FF$.  The weight functions $w_f(Q_1,Q_2)$ and
$w_g(M_V,Q_1,Q_2)$ are both positive and peaked around $Q_1 \sim Q_2
\sim 0.5$~GeV.  In the case of the vector meson dominance (VMD) form
factor, $f^{VMD}(M_V,Q_1,Q_2)$ vanishes, and only the term involving
$w_g(M_V,Q_1,Q_2)$ contributes.  The corresponding value obtained in
\cite{KN_01} for $a_{\mu}^{\mbox{\tiny{LbyL;}}\pi^0}\vert_{VMD}$ upon
numerical evaluation of the integral (\ref{2dInt}) differs only by its
overall sign from the results obtained by previous authors
\cite{HKS_95_96,BPP}. Now, in Ref. \cite{KN_01}, it has been shown
that (up to $\sim \ln M_V$ or constant terms)
\bea
&&
\lim_{M_V\to\infty} 
\int_0^{\infty}\!\!\!\!dQ_1\!\!\int_0^{\infty}\!\!\!\!dQ_2\,
g^{{\scriptscriptstyle VMD}}(M_V,Q_1,Q_2) w_{g}(M_V,Q_1,Q_2) 
\nonumber\\
&&
%\qquad \,=\,
\quad ={\cal C}\ln^2 M_V\,, % + \cdots 
\eea
with a constant ${\cal C}$ that numerically agrees with the result of
Eq. (\ref{Clog2}), including its {\it sign}. A global sign error in
the second contribution to (\ref{2dInt}) is thus excluded. On the
other hand, for a constant form factor, normalized by the WZW term,
$f^{WZW}(M_V,Q_1,Q_2) = [N_C / (12\pi^2 F_\pi)]^2$ and
$g^{WZW}(M_V,Q_1,Q_2)$ vanishes. In that case, the integral involving
$w_{f}(Q_1,Q_2)$ diverges.  In Ref. \cite{KN_01}, it was shown that
this divergence is of the form (up to $\sim \ln \Lambda$ or constant
terms)   
\bea
&&
\lim_{\Lambda\to\infty} 
\int_0^{\Lambda}\!\!\!\!dQ_1\!\!\int_0^{\Lambda}\!\!\!\!dQ_2\,
f^{{\scriptscriptstyle WZW}}(M_V,Q_1,Q_2) w_{f}(Q_1,Q_2)
\nonumber\\
&&
%\qquad \,=\, 
\quad ={\cal C}\ln^2 \Lambda\,, % + \cdots , 
\eea
where, again, the constant ${\cal C}$ was found \cite{KN_01} to be {\it
positive} and in agreement with Eq. (\ref{Clog2}). We thus conclude
that the overall sign of the first term in Eq. (\ref{2dInt}) also has
to be correct. 

Therefore, we expect that $a_{\mu}^{\mbox{\tiny{LbyL;}}\pi^0}$ is a
{\it positive} quantity, at least for the VMD type of parametrizations
of the form factor $\FF$. This result was also found to hold for a
wider class of form factors studied in Ref. \cite{KN_01}.  
The implications of these observations for the comparison between 
theory and experiment  as far as the
anomalous magnetic moment of the muon is concerned have already been 
discussed in Ref. \cite{KN_01}. In particular,  
the difference between the present experimental value of $a_{\mu}$ 
\cite{BNL2} and the standard model is reduced from 2.6$\sigma$
to 1.5$\sigma$. 
The extension of the present effective 
field theory analysis to the other contributions from hadronic 
light-by-light scattering to the muon $g-2$ will be discussed elsewhere.

%\vspace*{0.3cm} 
This work has been supported in part by Schweizerischer Nationalfonds
and by TMR, EC Contract No. ERBFMRX-CT980169 (EURODAPHNE).

\end{document}